\documentclass[aps,prfluids,notitlepage,longbibliography]{revtex4-1}  

\usepackage[utf8]{inputenc}
\usepackage{epsf,graphicx,color}
\usepackage{multirow}
\usepackage{amsmath,gensymb,amssymb}
\usepackage{siunitx}
\usepackage[bottom]{footmisc}
\usepackage{lipsum}
\usepackage{dcolumn}
\usepackage{bm}

\usepackage{hyperref}
\hypersetup{
citecolor=magenta
breaklinks=true, 
urlcolor= blue, 
linkcolor= blue, 
bookmarksopen=true, 
}

\usepackage{natbib}

\usepackage{color}
\definecolor{linkcolor}{rgb}{0,0,0.6}

\begin{document}

\title{Three-dimensional turbulence generated homogeneously by magnetic particles}%

\author{A. Cazaubiel}%
\author{J.-B. Gorce}%
\author{J.-C. Bacri}%
\author{M. Berhanu}%
\author{C. Laroche}%
\author{E. Falcon}%
\email[E-mail: ]{eric.falcon@u-paris.fr}
\affiliation{Universit\'e de Paris, MSC, UMR 7057 CNRS, F-75 013 Paris, France}
\begin{abstract}
Three-dimensional turbulence is usually studied experimentally by using a spatially localized forcing at large scales (e.g. via rotating blades or oscillating grids), often in a deterministic way. Here, we report an original technique where the fluid is forced in volume, randomly in space and time, using small magnetic particles remotely driven. Such a forcing generates almost no mean flow and is closer to those of direct numerical simulations of isotropic homogeneous turbulence. We compute the energy spectra and structure functions using local and spatiotemporal flow velocity measurements. The energy dissipation rate is also evaluated consistently in five different ways. Our experimental results confirm the stationary, homogeneous and isotropic features of such turbulence, and in particular the Tennekes' model for which the Tennekes' constant is experimentally estimated.  
\end{abstract}
\maketitle

\section{Introduction}
Turbulence concerns swirling motions of fluids occurring irregularly in space and time. This phenomenon occurs in most geophysical or astrophysical flows, as well as in many industrial processes \cite{Davidson,GaltierBook}. However, attempts to find analytical solutions to the forced Navier-Stokes equations in a turbulent regime still remain unsuccessful. Three-dimensional turbulence is thus mainly described phenomenologically using dimensional and similarity arguments assuming notably homogeneity, isotropy and statistical stationarity \cite{K41,K41b,Pope,Davidson,Frisch}. For a long time, 3D turbulence experiments consisted of uniform grids of bars in a wind tunnel (freely decaying turbulence) to get closer to ideal isotropic and homogeneous turbulence \cite{Pope,Davidson}. Nowadays, most laboratory experiments on 3D stationary turbulence are performed in a closed container where energy is injected from a boundary of the container, at large scales and often in a deterministic way, such as oscillating grids \cite{Srdic1996,KitPoF1995,DeSilvaPoF1994,AlHomoudEFM2007}, counter-rotating disks (von K\'arm\'an flow) \cite{Douady91}, several fans \cite{Birouk2003} or propellers \cite{Guala2008}, or multiple jets \cite{Hwang2004,VarianoJFM2008,JohnsonCowenJFM2018}. In contrast, direct numerical simulations of 3D turbulence use a forcing in volume either in spectral space \cite{Eswaran1988}, or more recently in physical space \cite{Lundgren03}. To be able to experimentally force turbulence in the whole volume of a container (if possible randomly in time and space) is a challenge that has never been achieved to our knowledge, and would thus lead to a better comparison with direct numerical simulations \cite{Iyer2021}. 



Here, we present an original forcing technique where the fluid is forced in volume randomly in space and time, by using small magnetic particles remotely driven. An external oscillating magnetic field drives stochastic rotation of each magnetic particle, whereas the collisions between particles or with the container boundaries lead to erratic translational motions. Such a forcing within the bulk favors the statistical homogeneity of the velocity field with nearly no mean flow. The measured energy spectra, structure functions and energy dissipation rate (evaluated consistently in five different ways) confirm the stationary, homogeneous and isotropy features of such generated turbulence that could be easily implemented in different domains. 

Beyond its implementation to measure global dissipated power in 3D turbulence \cite{FalconPRF17}, this forcing mechanism can be also easily used in other systems as in soft matter to study a 3D granular ``gas" in air (showing several major differences with a boundary-driven system)\cite{FalconEPL13}.  Furthermore, colloidal magnetic spinners on a fluid surface, as well as active (self-propelled) swimmers, can generate flow reminiscent of 2D turbulence~\cite{KokotPNAS2017,BourgoinPRX20}. 



\section{Theoretical backgrounds}
For large enough Reynolds numbers and 3D stationary, homogeneous, and isotropic turbulence, the energy spectrum is predicted dimensionally as $E(k)=\mathcal{C}\epsilon^{2/3}k^{–5/3}$ \cite{K41} with $\epsilon$ the energy dissipation rate per unit mass and $k$ the Fourier spatial scale, and $\mathcal{C}\approx 1.6$ the Kolmogorov constant measured experimentally  \cite{Pope,SaddoughiJFM94}. $\epsilon$ also represents the mean flux of kinetic energy cascading from the large (forcing) scale to the small (dissipative) scale. This energy transfer through this inertial range is due to nonlinearity. The unidimensional (transverse and longitudinal) energy spectra are proportional theoretically as $E_{\perp}(k_x)=4/3E_{\parallel}(k_x)$ with $E_{\parallel}(k_x)=C\epsilon^{2/3}k_x^{–5/3}$ and $C=18\mathcal{C}/55$ \cite{Pope,SaddoughiJFM94}. The second-order moment of the velocity increments at a distance $r$ (or structure function) $\mathcal{S}_2(r) \equiv \langle [v(x+r)-v(x)]^2 \rangle$ is dimensionally predicted as $\mathcal{S}_2(r)=C_2\epsilon^{2/3} r^{2/3}$ \cite{K41}, $x$ is a spatial coordinate, and $C_2\approx 2.0$ is an experimentally measured constant~\cite{Pope}. The third-order structure function is analytically derived as $\mathcal{S}_3(r)=-4/5\epsilon r$ (the only exact result known for turbulence) called Kolmogorov's $4/5$ law \cite{K41b}. Finally, intermittency occurs if the structure functions of order $p$, $\mathcal{S}_p(r) \equiv \langle [v(x+r)-v(x)]^p \rangle$, scales as $r^{\zeta_p}$ with a nonlinear dependence of $\zeta_p$ with $p$ \cite{K62}, instead of $\zeta_p=p/3$~\cite{K41}. For finite Reynolds numbers, the previous laws have several corrections \cite{Pope,Karman38}.   

\begin{figure}
\includegraphics[width=0.5\linewidth]{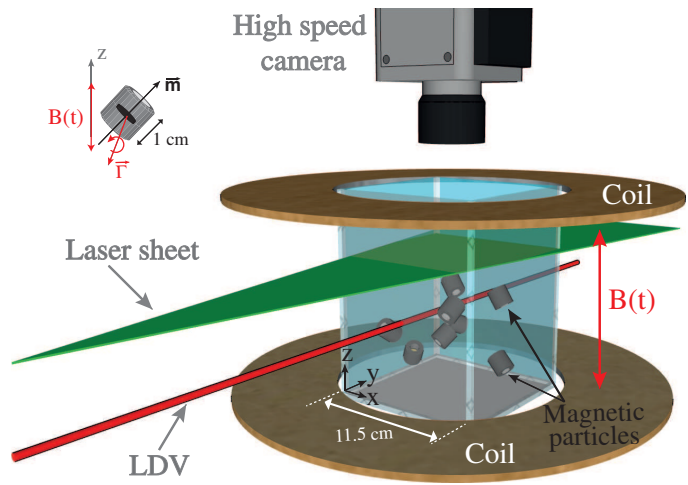} 
\caption{Experimental setup showing the 3D container of fluid and the encapsulated magnets together with PIV and LDV measurements. Top left: enlargement of a magnetic particle. A vertical oscillating magnetic field $B(t)$ drives time-dependent rotations of each magnetic particle by applying a torque $\vec{\Gamma}$ over its magnetic moment $\vec{\mathsf{m}}$.}
\label{fig1}
\end{figure}

\section{Experimental setup}
The experimental setup is shown in Fig. \ref{fig1}. A Plexiglas square-section container of length $L=11.5$ cm and height $h=9$ cm, is filled with distilled water. $N$ home-made magnetic particles are put within the container ($N\in[1, 60]$). Each particle is made of a cylindrical permanent  neodymium magnet (NdFeB, N52, 0.5 cm in diameter, 0.2 cm in thickness) encased and axially aligned in a cylindrical Plexiglas shell (1 cm in outer diameter and 1 cm long) to strongly reduce dipolar interaction between particles \cite{FalconPRF17}. The container is sealed with a transparent lid and sits between two Helmholtz coils powered by a sinusoidal current of amplitude $I\in[0, 9]$ A and frequency $F\in[0, 50]$ Hz. A vertical oscillating magnetic field $B(t)=B\sin{(2\pi F t)}$ is thus generated with an amplitude $B\in[0, 207]$ G measured with a gaussmeter (FW Bell). $B$ is spatially homogeneous in the container volume with a 5\% accuracy. The AC magnetic field transfers angular momentum into each particle which is converted into linear momentum during collisions, leading to erratic translational and rotational motions of the particles (see \cite{FalconPRF17} for details). The fluid is thus forced homogeneously in volume, and randomly in both space and time. The fluid velocity is measured in a single point over time by nonintrusive Laser Doppler Velocimetry (LDV Dantec Flow Explorer 1D) to access to its frequency spectrum. The fluid velocity field is measured in a horizontal $xy$ plane (11 $\times$ 9 cm$^2$) over time by Particle Image Velocimetry (PIV) \cite{RaffelPIV}, in particular to access the wavenumber spectrum and structure functions. The fluid flow is visualized using Polyamide fluid tracers (50 $\mathrm{\mu}$m) illuminated by a horizontal laser sheet. A high-resolution video camera (Phantom V10, 2400 $\times$ 1800 pixel$^2$ at 200 fps), located on the top of the fluid container, records the motion of the fluid tracers. The spatial resolution is 0.8 mm (i.e., spacing between adjacent velocity field vectors). Note that less than 3\% of the acquired images are discarded and correspond to rare events of a magnetic particle passing through the laser sheet. This leads to experiments for PIV with a lower $N$ and at lower fluid RMS velocity ($\sigma_u \leq 4$ cm/s) than for LDV ($\sigma_u \leq 18$ cm/s). For most of the results presented below, the volume fraction is 0.7\% (corresponding to $N=10$). 

Figure~\ref{fig1b} shows the typical fluid motions characteristic of a turbulent flow (see also movies in the Supplemental Material \cite{SuppMat}). 
Strong spatial and temporal fluctuations of the flow are observed over various scales, together with eddies. We will characterize hereafter the properties of such turbulent flow generated by this novel forcing. We will also verify if a self-similar energy transfer  through the scales occurs by nonlinearity.

\begin{figure}
\includegraphics[width=0.5\linewidth]{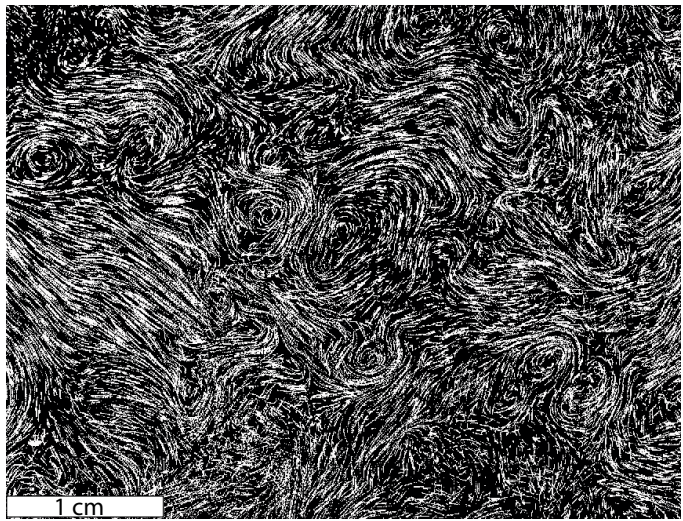} 
\caption{Fluid tracer trajectories within the laser sheet followed over 10 consecutive images (0.05 s). Forcing parameters: $N=10$, $B=115$ G, and $F=50$ Hz. $\sigma_u=2.3$ cm/s.}
\label{fig1b}
\end{figure}

\section{Homogeneity, isotropy and level of turbulence with control parameters}
The longitudinal and transverse horizontal fluid velocities at a location $x$ are defined as $u(x,t)$ and $v(x,t)$, the vertical one is $w(x,t)$. Using PIV, we first check that the RMS fluctuating velocity $\sigma_u$ is well invariant by translation in the $xy$ plane, and by rotation of the latter, meaning thus that the velocity field is homogeneous and isotropic in the horizontal plane. The isotropy ratios are indeed $\sigma_u/\sigma_v=0.97 \pm 0.01$ and $\sigma_u/\sigma_w=0.87 \pm 0.01$. Moreover, the mean velocity $\langle u \rangle_{t,x}$ is found to be much smaller than the RMS fluctuations (i.e., $\langle u \rangle_{t,x} / \langle \sigma_u \rangle_{x} <$ 11\%) to be able to neglect afterward the mean flow (see the Supplemental Material \cite{SuppMat}).

Using single-point LDV measurements, we now focus on the scalings of the fluid velocity fluctuations with the forcing parameters (number of magnetic particles $N$, amplitude $B$ and frequency $F$ of the magnetic field). The fluid RMS velocity fluctuations $\sigma_u=\sqrt{\left\langle u^2 \right\rangle_t}$ are found to depend on the forcing parameters  as $\sigma_u \sim N^{1/2}B^{1/3}F^{1/3}$ (see the Supplemental Material \cite{SuppMat}). The magnetic particle velocity was previously found to scale as $V_p\sim N^0B^{1/3}F^{1/3}$ from the power budget between the injected power into the fluid by the magnetic particles and the power dissipated \cite{FalconPRF17}. The latter is mainly due to viscous dissipation by a turbulent translational drag on the particles and by inelastic collisions between particles (or with the container walls) \cite{FalconPRF17}. Assuming that the kinetic energy of the fluid is proportional to the particle ones $\sim NV_p^2$, the RMS fluid velocity scales indeed as $\sigma_u \sim (NV_p^2)^{1/2} \sim N^{1/2}B^{1/3}F^{1/3}$. 

  \begin{figure}
 \includegraphics[width=0.5\linewidth]{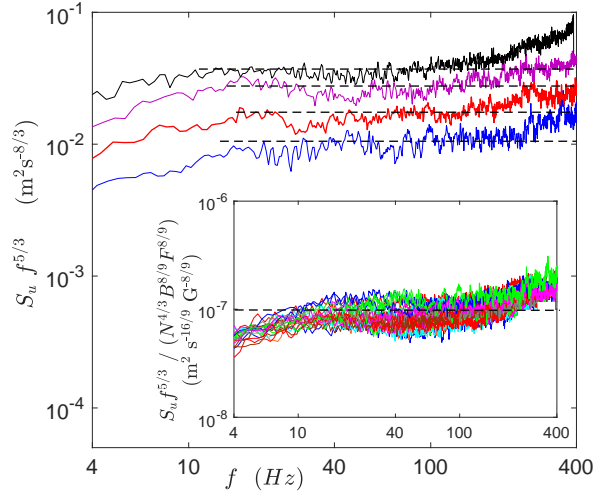} 
 \caption{Frequency power spectrum of the velocity $u(t)$ compensated by $f^{-5/3}$, $S_u(f) f^{5/3}$, for different $N$ from 10 (bottom) to 60 (top), with $F= 30$ Hz and $B$=161 G. Dashed lines correspond to the predictions (see text). Inset: compensated power spectra, $S_u(f) f^{5/3}$ rescaled by $N^{4/3}B^{8/9}F^{8/9}$ for various $N\in[10,60]$, $B\in[103,184]$ G, and $F\in[5, 55]$ Hz.}
 \label{Fig2}
  \end{figure}

\section{Frequency spectrum}
The power spectrum density $S_u(f)$ of the fluid velocity $u(t)$ measured by LDV is shown in Fig.~\ref{Fig2} and compensated by $f^{-5/3}$ for an increasing number $N$ of magnetic particles at fixed $B$ and $F$. The spectrum amplitude increased with $N$. More importantly, each spectrum follows a frequency power-law in $f^{-5/3}$ over more than one decade in frequency. 
For zero-mean velocity flows, Tennekes' model (large-scale advection of turbulent eddies) predicts the frequency spectrum to scale as $f^{-5/3}$~\cite{Tennekes}, as observed here. More precisely, one would expect $S(\omega)=\beta\epsilon^{2/3} q^{2/3}\omega^{-5/3}$ with  $\beta$ an empirical constant and $q\equiv \sqrt{(\sigma^2_u+\sigma^2_v+\sigma^2_w)}$~\cite{Tennekes}. Since $\epsilon \sim \sigma_u^3$ (see below), $S(\omega)$ has to scale as $\sigma_u^{8/3}$. We thus plot in the inset of Fig.~\ref{Fig2} the compensated spectra $S_u(f) f^{5/3}$ rescaled by $N^{4/3}B^{8/9}F^{8/9}$ for a large range of forcing parameters ($N$, $B$ and $F$). All rescaled spectra are well superimposed on a master curve with a plateau over more than one decade. As we confirm the Tennekes' model, we are then able to infer experimentally the Tennekes' constant from the compensated spectra and $\epsilon$ values. One finds $\beta=0.64\pm0.15$. This value confirms the assumed Tennekes' constant of the order of 1 \cite{Tennekes} and simulations leading to $\beta=0.82$ \cite{FungJFM2008}. Note that the rare previous experimental estimates (mainly on smaller inertial ranges) vary from $\beta=0.14$ \cite{JohnsonCowenJFM2018} (resp. 0.23~\cite{VarianoJFM2008}) using multiple jets forcing, without (resp. with) a free surface, to $\beta \in[0.48, 0.62]$ \cite{KitPoF1995} and 5.5 \cite{DeSilvaPoF1994} using oscillating grids forcing but without PIV measurements, or $\beta\in[0.28, 3.5]$ for low Reynolds number flows \cite{AlHomoudEFM2007}.

  \begin{figure}
\includegraphics[width=0.5\linewidth]{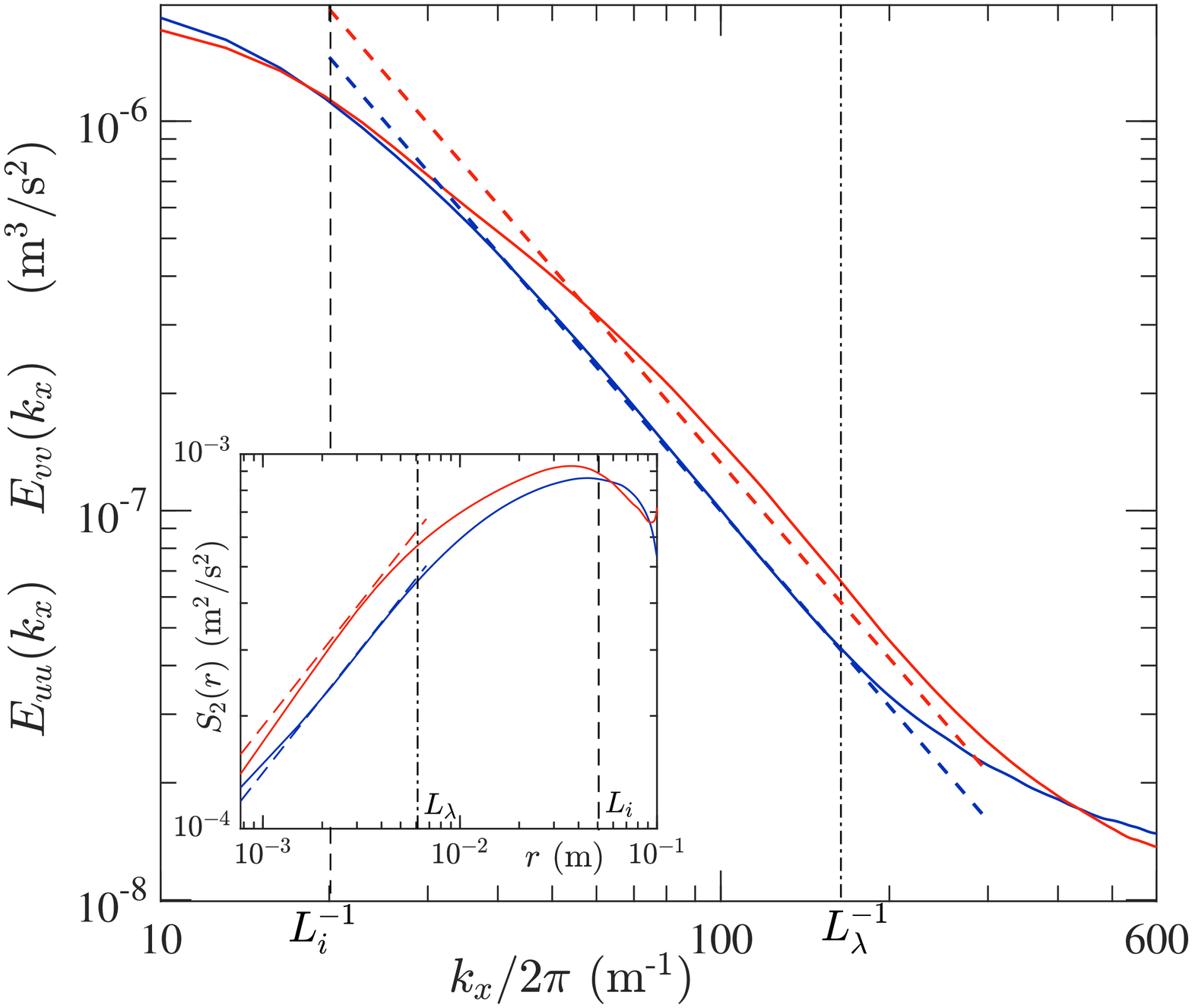} 
\caption{1D wavenumber power spectrum $E_{uu}(k_x)$ (in blue) and $E_{vv}(k_x)$ (red) in the $x$ direction of the velocity components $u$ and $v$. $N=10$, $B=115$ G and $F=50$ Hz. $\sigma_u= 2$ cm/s. Dashed lines: $k_x^{-5/3}$ prediction (blue one is adjusted and red one is inferred from the prediction $E_{vv}=\frac{4}{3}E_{uu}$). Vertical lines correspond to the length scales $L_i$ (dotted line) and $L_{\lambda}$ (dotted-dash line). Inset: Second-order structure functions $\mathcal{S}_2^{(u)}(r)$ (blue) and $\mathcal{S}_2^{(v)}(r)$ (red). Blue dashed line: best fit of $\mathcal{S}_2^{(u)}$ in $r^{2/3}$. The red dashed line is derived from the blue line using the relationship $\mathcal{S}_2^{(v)}=\frac{4}{3}\mathcal{S}_2^{(u)}$.}
\label{fig3}
\end{figure}
  
\section{Wavenumber spectrum and characteristic scales}
Using PIV, the 1D wavenumber power spectra (in the $x$ direction), $E_{uu}(k_x)$ and $E_{vv}(k_x)$, of the longitudinal and transverse components ($u$ and $v$) of the velocity field are shown in Fig.~\ref{fig3}. The longitudinal spectrum $E_{uu}(k_x)$ scales as $k_x^{-5/3}$ over a decade as expected from Kolmogorov's law $E_{uu} = C\epsilon^{2/3} k_x^{-5/3}$ \cite{K41}. We also observe that the transverse spectrum $E_{vv}(k_x)$ is proportional to the longitudinal one in agreement with $E_{vv}(k_x)=\frac{4}{3}E_{uu}(k_x)$ \cite{Pope} (see dashed lines in Fig.~\ref{fig3}). The degree of isotropy is thus comparable with that in DNS where the same equivalence between 1D spectra is found \cite{IyerPRE2015}.

The inertial scales of turbulence are located between the container size $L$ and the small dissipative Kolmogorov scale $\eta=(\nu^3/\epsilon)^{1/4}$ \cite{K41}. Here, one has $\eta \approx 0.2$ mm for a typical mean dissipation rate $\epsilon=10^{-3}$ m$^2$ s$^{-3}$ (see below), $\nu=10^{-6}$ m$^2$ s$^{-1}$ is the fluid kinematic viscosity. The integral length scale can not be accurately computed from the autocorrelation function of the velocity field since the container size $L$ is not eight times larger than the integral scale \cite{Pope,OneillProc04}.  We evaluate the integral scale $L_i=5$ cm from the abscissa of the maximum of $\mathcal{S}_2(r)$ (see inset of Fig.~\ref{fig3}), corresponding to roughly the beginning of the inertial range (see Fig.~\ref{fig3}). 
 The corresponding turbulent Reynolds number at $L_i$ thus reads $Re_{L_{i}}=\sigma_u L_i/ \nu \approx 10^3$, with $\sigma_u=2$ cm/s. The Taylor length scale is estimated as $L_{\lambda}\approx 6$ mm (well located between $L_i$ and $\eta$ - see Fig.~\ref{fig3}) using $L_{\lambda}=L_i\sqrt{15/Re_{L_{i}}}$~\cite{Davidson}. The corresponding Taylor Reynolds number is $Re_{\lambda}=\sigma_u L_{\lambda}/ \nu \approx 122$, a value of the same order of magnitude as the ones in boundary forced turbulence experiments \cite{Douady91,Birouk2003,Guala2008,Hwang2004}.   

\begin{figure}
\includegraphics[width=0.5\linewidth]{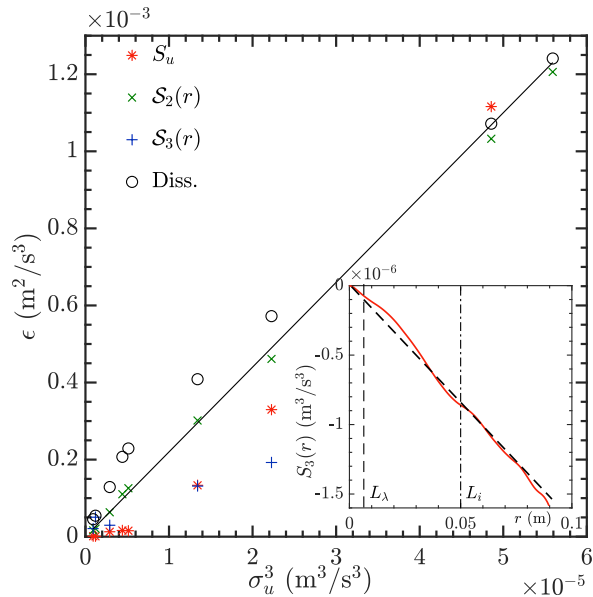} 
\caption{Different estimations of $\epsilon$ as a function of $\sigma_u^3$ inferred from ($\ast$) the wavenumber spectrum $E_{uu}$, ($\times$) the second-order structure function $\mathcal{S}_2(r)$,  ($+$) the third-order structure function $\mathcal{S}_3(r)$, and ($\circ$) the dissipation rate definition (see text). Solid line is the prediction $c\sigma_u^3/L_i$ with $c=1.1$. Inset: Third-order structure function $\mathcal{S}_3(r)$ (solid line). Dashed line is a linear fit.}
\label{Fig4}
\end{figure}

\section{Structure functions}
The structure functions of the velocity field are also computed from the PIV measurements. The inset of Fig. \ref{fig3} shows the second-order structure functions $\mathcal{S}_2(r)$ in the $x$ direction of the horizontal components of the velocity field $u$ and $v$. The structure functions $\mathcal{S}_2(r)$ are roughly proportional to $r^{2/3}$ in the inertial range, as expected by the $2/3$ Kolmogorov's law (see above) \cite{K41}. Moreover, as for the spectra, the transverse and longitudinal components are found proportional as $\mathcal{S}_2^{(v)}=\frac{4}{3}\mathcal{S}_2^{(u)}$ (see dashed lines) as expected theoretically. From $\mathcal{S}_2^{(u)}$ and $E_{uu}$, one can also infer the ratio of the $2/3$ law constant over the Kolmogorov's constant, $C_2/C=5.3\pm2.8$, not so far from previous experimental evaluations $\approx 4$ \cite{Pope}. The third-order structure function $\mathcal{S}_3^{(u)}$ of the longitudinal velocity field is also computed and shown in the inset of Fig.~\ref{Fig4}. $\mathcal{S}_3^{(u)}$ is found to decrease linearly with $r$ over one decade in the inertial range, in good agreement with the $4/5$ Kolmogorov's law~\cite{K41b} and with DNS~\cite{IyerPRF2020}. This corresponds to the negative asymmetry of the velocity fluctuation gradients quantified by the skewness $\mathcal{S}_3^{(u)}/(\mathcal{S}_2^{(u)})^{3/2}=-0.3\pm0.2$ close to the value inferred from the $4/5$ law, $-4/(5C_2)=-0.4$ \cite{Pope}. 


\section{Energy dissipation rate}
Finally, the mean energy dissipation rate $\epsilon$ is estimated in five different ways: (i) as $E_{uu}^{3/2} k_x^{5/2}/C^{3/2}$ from the experimental 1D wavenumber spectrum and the Kolmogorov's spectrum, (ii) as $[\mathcal{S}_2^{(u)}/C_2]^{3/2}/r$ from the experimental $\mathcal{S}_2^{(u)}$ and the $2/3$ law, (iii) as $-5\mathcal{S}_3^{(u)}/(4r)$ from the experimental $\mathcal{S}_3^{(u)}$ and the $4/5$ law, (iv) from its definition for isotropic turbulence $\epsilon\equiv15\nu \left\langle (\partial u_x/\partial x)^2\right\rangle$ \cite{Taylor35,Pope}, and (v) from dimensional analysis. These different estimations of $\epsilon$ are plotted in Fig. \ref{Fig4} as a function of $\sigma_u^3$. All $\epsilon$ values are found of the same order of magnitude at fixed $\sigma_u$, and are proportional to $\sigma_u^3$ regardless of the method used. Dimensional arguments estimate the dissipation rate from the velocity fluctuations as $\epsilon=c\sigma_u^3/L_i$  \cite{Taylor35,Pearson02} involving the integral scale $L_i$, and $c$ a constant of the order of unity \cite{Sreenivasan84,LohsePRL94}. Here, one finds $c=1.1$ close to the values found with a boundary forcing (grid turbulence) \cite{Pearson02,Wang2021}. These estimations of $\epsilon$ by five different methods are hardly obtained experimentally \cite{HoqueCES2015} and are found here to be all consistent as a consequence of the stationary, homogeneous and isotropic turbulence generated by this forcing in volume. Note that higher turbulence levels can be explored with this forcing (e.g., $\epsilon \sim 6\ 10^{-3}$ m$^2$/s$^{-3}$ for $\sigma_u \sim 0.18$ m/s measured with LDV).  

\section{Conclusion}
We developed an original technique to generate 3D turbulence by injecting energy in volume, randomly in time and space, by using small magnetic particles remotely driven. This forcing contrasts with previous ones in which a spatially localized forcing is applied at large-scale from a container boundary. We characterize the turbulence generated by this forcing in volume by local and spatiotemporal measurements of the fluid velocity. Almost no mean flow is involved, and all measured properties confirm the stationary, homogeneous and isotropic features of such turbulence. In particular, we confirm experimentally the Tennekes' model and resolve the disagreement between previously suggested value of the Tennekes' constant. Possible intermittency of such generated turbulence could be explored in the future \cite{BatchelorPRSA49}, as well as its Lagrangian properties \cite{ToschiARFM09}. Moreover, this forcing mechanism is closer to those of direct numerical simulations and is rather flexible (e.g. either random in space and time or random only in space or only in time). It appears very promising to study large-scale 3D turbulence (i.e. larger than the injection scale) and its possible description by statistical mechanics tools \cite{DallasPRL15}. It could be also applied to smart control of turbulence \cite{BuzzicottiPRL20}. Finally, this homogeneous forcing could be used to better explore geophysical- or astrophysical-like turbulent flows (rotating, stratified, or multiphase flows), and could provide a technological breakthrough in turbulent mixing.    



\begin{acknowledgments}
This work was supported by the French National Research Agency (ANR DYSTURB project No. ANR-17-CE30-0004), and by the Simons Foundation MPS N$^{\rm o}$651463-Wave Turbulence.
\end{acknowledgments}


\end{document}